\documentclass[draftclsnofoot, letterpaper, onecolumn, 12pt]{IEEEtran}

\usepackage{multirow}
\usepackage{booktabs}
\usepackage{amsfonts}
\usepackage{amsmath,cases}
\usepackage{amssymb}
\usepackage{graphicx}
\usepackage{cite}
\usepackage{epsfig}
\usepackage{url}
\usepackage{algorithm}
\usepackage{algorithmic}
\usepackage{epstopdf}
\usepackage{color}
\usepackage[tight]{subfigure}

\newcommand{\ds}{\displaystyle}
\newtheorem{myth}{Theorem}

\newcommand{\qed}{\mbox{}\hfill$\Box$ \vskip 3mm}

\newcommand{\la}{\langle}
\newcommand{\ra}{\rangle}
\newcommand{\Prf}{{\it Proof: \ }}

\newcommand{\clC}{{\cal C}}
\newcommand{\clL}{{\cal L}}

\newcommand{\diag}{{\sf diag}}
\newcommand{\Ex}{{\mathbb E}}

\begin{document}

\title{Joint Power Allocation and  Beamforming for Energy-Efficient  Two-Way Multi-Relay Communications
\thanks{This work was supported in part  by the Australian Research Council’s Discovery Projects under Project DP130104617, in part by the U.K. Royal Academy of Engineering Research Fellowship under Grant RF1415$\slash$14$\slash$22, and in part by
the U.S. National Science Foundation under Grants CCF-1420575 and CNS-1456793. }}
\author{Zhichao Sheng, Hoang D. Tuan, Trung Q. Duong and H. Vincent  Poor
\thanks{Zhichao Sheng and Hoang D. Tuan are with the Faculty of Engineering and Information Technology, University of Technology Sydney, Broadway, NSW 2007, Australia (email: kebon22@163.com, Tuan.Hoang@uts.edu.au)}
\thanks{Trung Q. Duong is with Queen's University Belfast, Belfast BT7 1NN, UK  (email: trung.q.duong@qub.ac.uk)}
\thanks{H. Vincent Poor is with the Department of Electrical Engineering, Princeton University, Princeton, NJ 08544, USA (e-mail: poor@princeton.edu)}
}

\maketitle
\vspace{-1.2cm}
\begin{abstract}
This paper considers the joint design of user power allocation and  relay beamforming in
relaying communications, in which multiple pairs of single-antenna users exchange information
with each other via multiple-antenna relays in two time slots. All users transmit their signals to the relays
in the first time slot while the relays broadcast the beamformed signals to all users in the second time slot.
The aim is to maximize the system's energy efficiency (EE) subject to
 quality-of-service (QoS) constraints in terms of exchange throughput requirements. The QoS constraints
are  nonconvex  with many nonlinear cross-terms, so finding a feasible point is already computationally challenging. The  sum throughput appears in the numerator while the total consumption power
appears in the denominator of the EE objective function. The former  is a  nonconcave function and the latter
is a nonconvex function, making fractional programming useless for EE optimization.
Nevertheless, efficient iterations of low complexity to obtain its optimized solutions are
developed. The performances of the multiple-user and multiple-relay networks under various scenarios are evaluated to show
the merit of the paper development.
\end{abstract}

\section{Introduction}\label{sec:intro}
Two-way relaying (TWR) \cite{Amarasuriya12,Chung12,Getal13} has been the focus of considerable research interest
in recent years due  to its potential in offering higher information exchange throughput for cognitive communications such as device-to-device (D2D) and machine-to-machine (M2M) communications \cite{AWM14,Liuetal15}.  Unlike
the conventional one-way relaying, which needs four time slots for information exchange between a pair of users
(UEs), TWR needs just two time slots for this exchange \cite{ZhangL09,Zeng11, Chung12}. In the first time slot, known as the multiple access (MAC) phase, all UEs simultaneously transmit their signals to the relays. In the second time slot, also known as the broadcast (BC) phase, the relays broadcast the beamformed signals to the all UEs.
Offering double fast communication, TWR obviously suffers from double
multi-channel interference as compared to one-way relaying  \cite{Chalise10,Phan13}. Both  optimal control of UEs' transmit power
and  TWR beamforming are thus very important in exploring the spectral efficiency of TWR.

There are various  scenarios of TWR considered in the literature. The most popular scenario is
single-antenna  relays serving a pair of single-antenna UEs \cite{Cheng11, Zeng11, Amirani12}.
The typical problems are to design the  TWR  weights to either maximize the  throughput
or minimize the relay power subject to signal-to-interference-plus-noise ratio (SINR) constraints at UEs.
A branch-and-bound (BB) algorithm of global optimization \cite{Tuybook} was used in \cite{Cheng11}
for sum throughput maximization. Its computational complexity is already very high in very low-dimensional problems.
Semi-definite relaxation  of high computational complexity
followed by bisection  used in \cite{Zeng11} works strictly under a single total relay power constraint.
Furthermore, the scenario of single-antenna relays serving multiple pairs of UEs was addressed in
\cite{Zhang12} by  a polyblock algorithm of global optimization
\cite{Tuybook} and in \cite{Zhang12} by local linearization based iteration. The mentioned semi-definite relaxation
was also used in \cite{Tao12,Wang12,KRVH12} in designing TWR beamformer for the scenario of
a multi-antenna relay serving multiple pairs of UEs. It should be realized that most of relay beamformer optimization
problems considered in all these works are not more difficult computationally than their one-way relaying counterparts,
which have been efficiently solved in \cite{Phan13}.

The fixed power allocation to UEs does not only miss the  opportunity of power distribution  within a network
but can also potentially increase interference to UEs of other networks \cite{Khandaker11, ChengPesavento12}. The joint design of UE power allocation and TWR weights for single antenna relays serving a pair of single antenna UEs
to maximize the minimum SINR was considered in \cite{Jing12}.  Under the strict assumption  that
the complex channel gains from UEs to the relays are the same as those from the relays to the UEs,
its design is divided into two steps. The first step is to optimize the beamformer weights with UEs'
fixed power by sequential second-order convex cone programming (SOCP).   The second step
performs an exhaustive grid search for UE power allocation. Joint optimization of UE
precoding and relay beamforming for a multi-antenna relay serving a pair of multi-antenna UEs
could be successfully addressed only recently in \cite{Raetal14}. An efficient computation for joint UE power
allocation and TWR beamforming to maximize the worst UE throughput  for multiple antenna relays
serving multiple pairs of single antenna UEs  was proposed
in \cite{Khetal13}. The reader is also referred to \cite{TNT17} for joint UE and relay power allocation
in MIMO OFDM system of one relay and one pair of UEs.

Meanwhile, the aforementioned classical spectral efficiency (SE) in terms of high  throughput is now only
one among multiple driving forces for the development of the next generation communication networks (5G) \cite{BJDO14}.
The energy consumption of communication systems has become sizable, raising environmental and economic
concerns \cite{Feetal11}. Particularly, the network energy efficiency (EE) in  terms of
the ratio of the sum throughput and the total power consumption, which counts not only the transmission power but also the
drain efficiency of power amplifiers, circuit power and other power factors in supporting the network's activities,
is comprehensively  pushed forward in 5G to address these concerns \cite{Buetal16}. EE in single-antenna TWR
has been considered in \cite{XLL15} for single-antenna OFDMA in assisting  multiple pairs of
single-antenna UEs and in \cite{ZH15} and \cite{WLTP16} for multi-antenna relays in assisting a pair of multi-antenna UEs. Again, the
main tool for computational solution in these works is semi-definite relaxation, which not only
significantly increases the problem dimension but also performs unpredictably \cite{Phan12}. Also, the resultant
Dinkelbach's iteration of fractional programming invokes  a logarithmic function optimization, which is convex but still
computationally difficult with no available algorithm of polynomial complexity.

The above analysis of the state-of-the-art TWR motivates us to consider the joint  design of single-antenna UE power
allocation and TWR beamformers in a TWR network to maximize its EE subject to UE  QoS constraints.
We emphasize that both sum throughput maximization (for spectral efficiency) and EE maximization are meaningful
only in the context of UE QoS satisfaction, without which they will cause the UE service discrimination.
Unfortunately, to our best knowledge such UE QoS constraints were not addressed whenever they are nonconvex \cite{Buetal16,XLL15,ZH15,WLTP16}. The nonconvexity of these QoS constraints implies that even finding their
feasible points is already computationally difficult. QoS constraints in terms of UEs' exchange throughput requirements
are much more difficult than that in terms of individual UE throughput requirements because the former cannot be
expressed in terms of individual SINR constraints  as the latter. To address the EE maximization problem, we first
develop a new computational method for UE exchange throughput requirement feasibility, which invokes only
simple convex quadratic optimization. A new path-following computational procedure for  computational solutions
of the EE maximization problem is then proposed.

The rest of the paper is organized as follows. Section
\ref{sec:Model} formulates  two optimization problems of EE maximization and UE QoS optimization.
Two  path-following algorithms are developed in Section \ref{sec:qos} and \ref{sec:ee} for their computation.
In contrast to fractional programming, these algorithms
invoke  only a simple convex quadratic optimization of low computational complexity at each iteration.
Section \ref{sec:Simulation} provides simulation results to verify the
performance of these algorithms. Finally, concluding remarks are
given in Section \ref{sec:Conclusion}

\emph{Notation.} Vectors and matrices are represented by boldfaced lowercase and uppercase, respectively.
 $\pmb{x}(n)$ is the $n$th entry of vector $\pmb{x}$, while $\pmb{X}(n,.)$ and $\pmb{X}(n,m)$  are the $n$th row and
  $(n,m)$th entry of a matrix $\pmb{X}$. $\langle \pmb{x},
\pmb{y}\rangle=\pmb{x}^H\pmb{y}$ is the inner product between
vectors $\pmb{x}$ and $\pmb{y}$. $||.||$ is either the Euclidean vector squared norm or
the Frobenius matrix squared norm, and $\mathbb{R}^N_+=\{\pmb{x}\in \mathbb{R}^N\ :\ \pmb{x}(n)\geq 0,\
n=1,2,\ldots,N\}$.  $\la \pmb{X} \ra$ is the trace of matrix $\pmb{X}$
 and $\diag[\pmb{X}_m]_{m=1,\ldots,M}$ is a block diagonal matrix with diagonal blocks $\pmb{X}_m$.
Lastly, $\pmb{x}\sim \mathcal{CN}(\bar{\pmb{x}},\pmb{R}_{\pmb{x}})$ means $\pmb{x}$ is a vector of Gaussian random variables with mean $\bar{\pmb{x}}$ and covariance $\pmb{R}_{\pmb{x}}$.
%

\section{Two-way Relay Networks with Multiple MIMO Relays and Multiple Single-Antenna Users}\label{sec:Model}
\begin{figure}[htb]
 \centering
 \centerline{\epsfig{figure=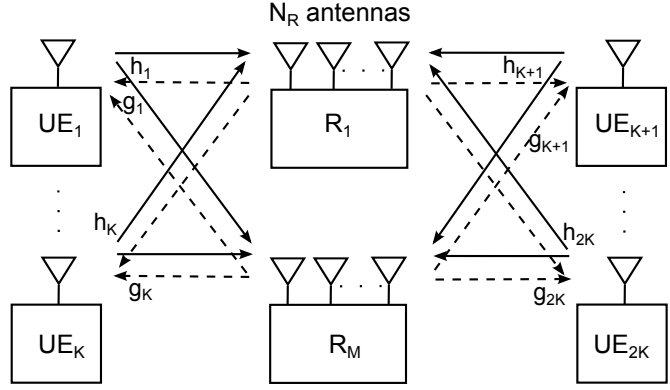,scale=1}}
\caption{Two-way relay networks with multiple single-antenna users and multiple multi-antenna relays.}
\label{fig:SystemModel}
\end{figure}

Fig. \ref{fig:SystemModel} illustrates a TWR network in which $K$ pairs of single-antenna UEs exchange information with each other. Namely the $k$th UE (UE $k$) and the $(K+k)$th UE (UE $K+k$), with $k=1,\ldots,K$, exchange information with
each other via
$M$ relays designated as relay $m$, $m=1,\dots,M$,  equipped with $N_R$ antennas.

Denote by $\pmb{s}=(s_1,\ldots,s_{2K})^T \in \mathbb{C}^{2K}$  the vector of  information symbols transmitted by the UEs, whose
entries are independent and have unit energy, i.e., $\Ex[\pmb{s}\pmb{s}^H]=\pmb{I}_{2K}$, where $\mathbf{I}_{2K}$ is
the identity matrix of size $(2K)\times (2K)$.
Let $\pmb{h}_{\ell,m}\in \mathbb{C}^{N_R}$ be the vector of channels from UE $\ell$ to  relay $m$. The received signal at  relay $m$ is
\begin{equation}\label{receivedr}
\pmb{r}_m=\sum_{\ell=1}^{2K} \sqrt{p_{\ell}} \pmb{h}_{\ell,m} s_{\ell}+\pmb{n}_{R,m},
\end{equation}
where {\color{black} $\pmb{n}_{R,m}\sim \mathcal{CN}(0,\sigma_R^2\pmb{I}_{N_R})$} is the background noise, and $\pmb{p}=(p_1,\ldots,p_{2K})^T \in \mathbb{R}_+^{2K}$ represents the powers allocated to the UEs.

Relay $m$ performs linear processing on the received signal by applying the beamforming matrix
$\pmb{W}_m \in \clC^{N_R \times N_R}$. The beamformed signals are
\begin{equation}
\pmb{r}_{m,b}=\pmb{W}_m\pmb{r}_m=\sum_{\ell=1}^{2K} \sqrt{p_{\ell}} \pmb{W}_m\pmb{h}_{\ell,m} s_{\ell}+{\color{black}
\pmb{W}_m}\pmb{n}_{R,m}, m=1,\dots, M.
\end{equation}
The transmit power at relay $m$ is calculated as
\[
 \Ex[||\pmb{r}_{m,b}||^2] = \ds \sum_{\ell=1}^{2K} p_{\ell} ||\pmb{W}_m\pmb{h}_{\ell,m}||^2+
   \sigma_R^2||\pmb{W}_m||^2.
\]
Relay $m$ transmits the beamformed signal $\pmb{r}_{m,b}$ to the UEs.
Let $\pmb{g}_{m,k}\in \mathbb{C}^{N_R}$ be the vector of channels from the relay $m$ to UE $k$.
The received signal at  UE $k$ is given by
\begin{eqnarray}
y_k&=&\ds\sum_{m=1}^M\pmb{g}_{m,k}^T\pmb{r}_{m,b}+n_k\nonumber\\
&=&\ds\sum_{m=1}^M\pmb{g}_{m,k}^T
(\sum_{\ell=1}^{2K} \sqrt{p_\ell}\pmb{W}_m\pmb{h}_{\ell,m} s_\ell+\pmb{W}_m\pmb{n}_{R,m})+n_k,\label{receivedu}
\end{eqnarray}
where  $n_{k}\sim \mathcal{CN}(0,\sigma_{k}^2)$ is the background noise, which can be written as
\begin{equation}\label{yk}
\begin{array}{lll}
y_k&=&\underbrace{\ds\sqrt{p_{\chi(k)}}\sum_{m=1}^M\pmb{g}_{m,k}^T
\pmb{W}_m\pmb{h}_{\chi(k),m} s_{\chi(k)}}_{\text{\sf desired signal}} +
\underbrace{\sqrt{p_k}\ds\sum_{m=1}^M\pmb{g}_{m,k}^T
\pmb{W}_m\pmb{h}_{k,m} s_{k}}_{\text{\sf self-interference}}\\
&&+ \underbrace{\ds\sum_{m=1}^M\pmb{g}_{m,k}^T(\sum_{\ell=1, \ell\neq k, {\chi(k)}}^{2K} \sqrt{p_\ell}\pmb{W}_m\pmb{h}_{\ell,m} s_\ell}_{\text{\sf inter-pair interference}}+
\pmb{W}_m\pmb{n}_{R,m})+n_k.
\end{array}
\end{equation}
In the above equation,  $(k, \chi(k))$ is a pair of UEs that exchange information with each other, so
\begin{equation}\label{pairm}
\chi(k)=\begin{cases}\begin{array}{lll}
K+k&\mbox{for}&k=1,\dots, K \cr
k-K&\mbox{for}& k=K+1,\dots, 2K
\end{array}
\end{cases}
\end{equation}
Assuming that the channel state information (CSI)  of the forward and backward channels  and the beamforming matrices
is available, UE $k$  effectively subtracts  the self-interference term in (\ref{yk}) to have the SINR:
\begin{eqnarray}
\gamma_{k}(\pmb{p}, \pmb{W})&=&\ds \frac{p_{\chi(k)}\left|\ds\sum_{m=1}^M\pmb{f}_{m,k}^H
\pmb{W}_m\pmb{h}_{\chi(k),m}\right|^2}{
\ds\sum_{\ell=1, \ell \neq k,\chi(k)}^{2K} p_{\ell}\left|\sum_{m=1}^M\pmb{f}_{m,k}^H \pmb{W}_m\pmb{h}_{\ell,m}\right|^2
+\sigma_R^2\sum_{m=1}^M||\pmb{f}_{m,k}^H \pmb{W}_m||^2+\sigma^2_{k}},
\label{SINR}
\end{eqnarray}
where {\color{black}$\pmb{f}_{m,k}=\pmb{g}_{m,k}^*$, so $\pmb{f}_{m,k}^H=\pmb{g}_{m,k}^T$}. \\
Under the definitions
\begin{equation}\label{ndef1}
\begin{array}{c}
{\color{black} \clL_{k,\ell}(\pmb{W})\triangleq\ds\sum_{m=1}^M\pmb{f}_{m,k}^H \pmb{W}_m\pmb{h}_{\ell,m},} \\
\clL_k(\pmb{W})\triangleq \begin{bmatrix}\pmb{f}_{1,k}^H \pmb{W}_1&\pmb{f}_{2,k}^H \pmb{W}_2&...&\pmb{f}_{M,k}^H \pmb{W}_M
\end{bmatrix},
\end{array}
\end{equation}
it follows that
\begin{eqnarray}
\gamma_{k}(\pmb{p}, \pmb{W})&=&\ds p_{\chi(k)}|\clL_{k,\chi(k)}(\pmb{W})|^2/
[\ds\sum_{\ell=1, \ell \neq k,\chi(k)}^{2K} p_{\ell}|\clL_{k,\ell}(\pmb{W})|^2+\sigma_R^2||\clL_k(\pmb{W})||^2 +\sigma^2_{k}].
\label{SINR.e}
\end{eqnarray}
In TWR networks, the UEs exchange information in bi-direction fashion in one time slot.
Thus, the  throughput at the $k$th UE pair
is defined by the following function of beamforming matrix $\pmb{W}$  and power allocation vector $\pmb{p}$:
\begin{eqnarray}
R_k(\pmb{p}, \pmb{W})&=& \ln ( 1+\gamma_{k}(\pmb{p}, \pmb{W}))+\ln (1+\gamma_{K+k}(\pmb{p}, \pmb{W})),  \quad k=1,\ldots,K.\label{Ratek}
\end{eqnarray}
Accordingly, the problem of maximizing the network EE subject to UE QoS constraints is
formulated as:
\begin{subequations}\label{e1}
\begin{eqnarray}
&\ds\max_{\pmb{W}\in\mathbb{C}^{N\times N}, \pmb{p}\in \mathbb{R}_+^{2K}} &F_{EE}(\mathbf{w},\mathbf{p})\triangleq
 \ds  {\color{black} \frac{ \sum_{k=1}^K\left[\ln (1+ \gamma_{k}(\pmb{p}, \pmb{W}))+\ln (1+\gamma_{K+k}(\pmb{p}, \pmb{W}))\right]}{\zeta(P^U_{\rm sum}(\mathbf{p})+P^R_{\rm sum}(\mathbf{p},
\mathbf{W}))+MP^{\rm R}+2KP^{\rm U}}} \label{e1a}\\
& \mbox{subject to} & 0\leq p_k \leq P_k^{U,\max}, \quad k=1,\ldots,2K, \label{IndiUserCon}\\
&&\sum_{k=1}^{2K} p_k \leq P_{\mathrm{sum}}^{U,\max}, \label{SumUserCon}\\
&&\ds \sum_{\ell=1}^{2K} p_{\ell}||\pmb{W}_m\pmb{h}_{\ell,m}||^2+
   \sigma_R^2||\pmb{W}_m||^2 \leq P_m^{A,\max}, m=1,...,M, \label{IndiRelayCon}\\
&&\ds\sum_{m=1}^M(\ds \sum_{{\color{black}\ell}=1}^{2K} p_{\ell}||\pmb{W}_m\pmb{h}_{\ell,m}||^2+
   \sigma_R^2||\pmb{W}_m||^2)\leq P_{\mathrm{sum}}^{R,\max}, \label{SumRelayCon}\\
&&R_k(\pmb{p}, \pmb{W})\geq r_{k}, k=1,\cdots, K,\label{e1c}
\end{eqnarray}
\end{subequations}
where $\zeta$, $P^{\rm R}$ and $P^{\rm U}$ are the reciprocal of drain efficiency of power amplifier,
the circuit powers of the relay and UE, respectively. {\color{black} $P^{\rm R}=N_R P_r$ and $P_r$ is the circuit power for each antenna in relay. } (\ref{e1c}) is the exchange throughput QoS requirement for each pair of UEs.
Constraints (\ref{IndiUserCon}) and (\ref{IndiRelayCon}) cap the transmit power of each UE $k$ and each
relay $m$ at predefined values $P_k^{U,\max}$ and $P_m^{A,\max}$, respectively. On the other hand, constraints
(\ref{SumUserCon}) and (\ref{SumRelayCon}) ensure that  the total transmit power of UEs and the total transmit
power of the relays not exceed the allowed power budgets $P_{\mathrm{sum}}^{U,\max}$ and $P_{\mathrm{sum}}^{R,\max}$,
respectively.

Note  that \eqref{e1} is a very difficult nonconvex optimization problem because
the power constraints \eqref{IndiRelayCon} and \eqref{SumRelayCon}, the exchange throughput QoS
constraints (\ref{e1c}), and the objective function \eqref{e1a}  are  nonconvex.
Moreover, the exchange throughput QoS constraints (\ref{e1c}) are much harder than the
typical individual throughput QoS constraints
\begin{equation}\label{indi1}
\ln(1+\gamma_k(\pmb{p},\pmb{W}))\geq \tilde{r}_{\min}, k=1,\dots, 2K,
\end{equation}
which are equivalent to the computationally easier SINR constraints
\begin{equation}\label{indi2}
\gamma_k(\pmb{p},\pmb{W})\geq e^{\tilde{r}_{\min}}-1, k=1,\dots, 2K.
\end{equation}
Note that even finding  feasible point of the EE problem (\ref{e1}) is already difficult as
it must be based on the following UEs QoS optimization problem
\begin{subequations}\label{MaxMinPair2}
\begin{eqnarray}
&\ds\max_{\pmb{W}\in\mathbb{C}^{N\times N}, \pmb{p}\in \mathbb{R}_+^{2K}} \ \varphi(\pmb{p},\pmb{W})\triangleq
\ds \min_{k=1,\ldots ,K}\ \left[\ln (1+ \gamma_{k}(\pmb{p}, \pmb{W}))^{1/r_k}+\ln (1+\gamma_{K+k}(\pmb{p}, \pmb{W}))^{1/r_k}\right] &\label{MaxMinPairObj2}\\
& \mbox{subject to} \quad \eqref{IndiUserCon}, \eqref{SumUserCon}, \eqref{IndiRelayCon}, \eqref{SumRelayCon},&\label{MaxMinPair2b}
\end{eqnarray}
\end{subequations}
which is still highly nonconvex because
the objective function in (\ref{MaxMinPairObj2}) is nonsmooth and nonconcave while the joint power constraints
\eqref{IndiRelayCon} and \eqref{SumRelayCon}
in (\ref{MaxMinPair2b}) are nonconvex. Only a particular case of $r_k\equiv r_{\min}$ was addressed in
\cite{Khetal13}, under which  (\ref{MaxMinPair2}) is then equivalent to the SINR multiplicative maximization
\begin{subequations}\label{multiplicative}
\begin{eqnarray}
&\ds\max_{\pmb{W}\in\mathbb{C}^{N\times N}, \pmb{p}\in \mathbb{R}_+^{2K}} &
\ds \min_{k=1,\ldots ,K}\  [(1+\gamma_{k}(\pmb{p}, \pmb{W}))(1+\gamma_{K+k}(\pmb{p}, \pmb{W})] \label{multiplicativea}\\
& \mbox{subject to} & \eqref{IndiUserCon}, \eqref{SumUserCon}, \eqref{IndiRelayCon}, \eqref{SumRelayCon},\label{multiplicativeb}
\end{eqnarray}
\end{subequations}
which can be solve by d.c. (\underline{d}ifference of two \underline{c}onvex functions) iterations \cite{KTN12}.
To the authors' best knowledge there is no available computation for (\ref{MaxMinPair2}) in general. The next
section is devoted to its solution.
\section{Maximin exchange throughput optimization}\label{sec:qos}
To address (\ref{MaxMinPair2}), our first major step is to transform the nonconvex constraints \eqref{IndiRelayCon} and \eqref{SumRelayCon} to convex ones through variable change as follows.\\
Following \cite{Khetal13}, make the variable change
\begin{equation}\label{betak}
\beta_k=\frac{1}{p_k^2} \geq 0, \quad k=1,\ldots,2K.
\end{equation}
For $\alpha>0$ and $\beta>0$, define the functions
\begin{equation}\label{Psi}
\begin{array}{c}
\Psi_{k,\ell}(\pmb{W}, \alpha, \beta)\triangleq\ds\frac{|\clL_{k,\ell}(\pmb{W})|^2}{\sqrt{\alpha\beta}}, \quad k,\ell=1,\ldots,2K,\\
\Upsilon_{k}(\pmb{W}, \alpha)\triangleq\ds\frac{||\clL_k(\pmb{W})||^2}{\sqrt{\alpha}}, \quad k=1,\ldots,2K,\\
\Phi_{\ell,m}(\pmb{W}_m, \alpha,\beta)\triangleq\ds\frac{||\pmb{h}_{\ell,m}^H\pmb{W}_m||^2 }{\sqrt{\alpha\beta}}, \quad \ell=1,\ldots,2K; m=1, \ldots, M.
\end{array}
\end{equation}
which are convex \cite{DM08}.
\begin{myth}\label{basic}
The optimization problem \eqref{MaxMinPair2} can be equivalently rewritten as
\begin{subequations}\label{MaxMinPair3}
\begin{eqnarray}
&\ds\max_{\pmb{W}\in\mathbb{C}^{N\times N}, \pmb{\alpha}\in \mathbb{R}_+^{2K}, \pmb{\beta}\in \mathbb{R}_+^{2K}} f(\pmb{W},\pmb{\alpha},
\pmb{\beta})\triangleq
\ds \min_{k=1,\ldots ,K}\ \frac{1}{r_k}
\left[\ln (1+\frac{|\clL_{k,K+k}(\pmb{W})|^2}
{\sqrt{\alpha_{k}\beta_{K+k}}})\right.\nonumber\\
&\left.\hspace*{8cm}+\ln (1+\ds\frac{|\clL_{K+k,k}(\pmb{W})|^2}
{\sqrt{\alpha_{K+k}\beta_{k}}}) \right] & \label{MaxMinPairObj3}\\
& \mbox{subject to} \quad \ds \sum_{\ell=1, \ell \neq k,\chi(k)}^{2K} \Psi_{k,\ell}(\pmb{W}, \alpha_k, \beta_{\ell})
+\sigma_R^2 \Upsilon_k(\pmb{W},\alpha_k)+{\color{black}\sigma_k^2\frac{1}{\sqrt{\alpha}_k}} \leq 1, \ k=1,\ldots,2K & \label{DemoniCon3}\\
& \beta_k \geq \frac{1}{(P_k^{U,\max})^2}, \quad k=1,\ldots,2K, & \label{UserCon3}\\
& P_{\mathrm{sum}}^{U}(\pmb{\beta}):=\ds \sum_{k=1}^{2K} \frac{1}{\sqrt{\beta_k}} \leq P_{\mathrm{sum}}^{U,\max}, & \label{TotalUserCon3}\\
& \ds \sum_{\ell=1}^{2K}\Phi_{\ell,m}(\pmb{W}_m,1,\beta_{\ell}) +\sigma_R^2||\pmb{W}_m||^2  \leq P_m^{A,\max},  m=1,\ldots,M, \label{RelayCon3}\\
& \ds\sum_{m=1}^M[\ds \sum_{\ell=1}^{2K}\Phi_{\ell,m}(\pmb{W}_m,1,\beta_{\ell}) +\sigma_R^2||\pmb{W}_m||^2 ]
\leq P_{\mathrm{sum}}^{R,\max}. & \label{TotalRelayCon3}
\end{eqnarray}
\end{subequations}
\end{myth}
\Prf One can see that $\gamma_{k}(\pmb{p}, \pmb{W})$ in (\ref{SINR.e}) is expressed in terms of functions in (\ref{Psi}) as
\begin{eqnarray}
\gamma_{k}(\pmb{p}, \pmb{W})&=&\ds\frac{ |\clL_{k,\chi(k)}(\pmb{W})|^2}{\frac{1}{p_{\chi(k)}}
[\ds\sum_{\ell=1, \ell \neq k,\chi(k)}^{2K}\Psi_{k,\ell}(\pmb{W},1,1/p_{\ell}^2)
+\sigma_R^2\Upsilon_k(\pmb{W},1)+\sigma^2_{k}]}\label{SINR.ea}
\end{eqnarray}
which is
\begin{equation}
\ds\frac{|\clL_{k,\chi(k)}(\pmb{W})|^2}
{\sqrt{\alpha_{k}\beta_{\chi(k)}}}\label{eq1}
\end{equation}
for
\begin{equation}\label{eq2}
\beta_k=1/p_k^2, k=1,\dots, 2K
\end{equation}
and
\begin{equation}\label{eq3}
\alpha_k=\ds\sum_{\ell=1, \ell \neq k,\chi(k)}^{2K}\Psi_{k,\ell}(\pmb{W},1,\beta_{\ell})
+\sigma_R^2\Upsilon_k(\pmb{W},1)+\sigma^2_{k}.
\end{equation}
Therefore,
\[
\varphi(\mathbf{p},\mathbf{W})=f(\mathbf{W},\pmb{\alpha},\pmb{\beta})
\]
for $\pmb{\alpha}$ and $\pmb{\beta}$ defined by (\ref{eq2})-(\ref{eq3}), which  is also feasible for  (\ref{MaxMinPair3})
whenever  $(\mathbf{p},\mathbf{W})$ is feasible for  (\ref{MaxMinPair2}). We thus have proved that
\begin{equation}\label{eq4}
\max\ (\ref{MaxMinPair2}) \leq  \max\ (\ref{MaxMinPair3}).
\end{equation}
Note that (\ref{DemoniCon3}) is the same as
\begin{equation}\label{eq5}
\ds\sum_{\ell=1, \ell \neq k,\chi(k)}^{2K}\Psi_{k,\ell}(\pmb{W},1,\beta_{\ell})
+\sigma_R^2\Upsilon_k(\pmb{W},1)+\sigma^2_{k}\leq \sqrt{\alpha_k}.
\end{equation}
The point $(\mathbf{p},\mathbf{W})$ with $p_k=1/\sqrt{\beta}_k$, $k=1,\dots, 2K$ is feasible for (\ref{MaxMinPair2})
whenever $(\mathbf{W},\pmb{\alpha}, \pmb{\beta})$ is feasible for (\ref{MaxMinPair3}). Using (\ref{SINR.ea}) and (\ref{eq5}),
one can see
\[
f(\pmb{W},\pmb{\alpha},\pmb{\beta})\leq \varphi(\pmb{p},\pmb{W}),
\]
implying
\[
\max\ (\ref{MaxMinPair3}) \leq  \max\ (\ref{MaxMinPair2}).
\]
The last inequality together with (\ref{eq4}) yield
\[
\max\ (\ref{MaxMinPair3}) \leq  \max\ (\ref{MaxMinPair2}),
\]
completing the proof of Theorem \ref{basic}.\qed

The benefit of expressing (\ref{MaxMinPair2}) by (\ref{MaxMinPair3})  is that all constraints in the latter
are convex so the computational difficulty is concentrated in its objective function, which is lower bounded by
a concave function based on the following result.

\begin{myth}\label{cth1} At $(\pmb{W}^{(\kappa)},\pmb{\alpha}^{(\kappa)},\pmb{\beta}^{(\kappa)})$ it
is true that
\begin{eqnarray}
\ln (1+\frac{|\clL_{k,\chi(k)}(\pmb{W})|^2}
{\sqrt{\alpha_{k}\beta_{\chi(k)}}})&\geq&f^{(\kappa)}_{k,\chi(k)}(\pmb{W},\alpha_{k},\beta_{\chi(k)})\nonumber\\
&\triangleq&a_{k,\chi(k)}^{(\kappa)}-b_{k,\chi(k)}^{(\kappa)}\sqrt{\alpha_{k}^{(\kappa)}\beta_{\chi(k)}^{(\kappa)}}
[2\Re\{\clL_{k,\chi(k)}(\pmb{W})(\clL_{k,\chi(k)}(\pmb{W}^{(\kappa)}))^*\}\nonumber\\
&&-\ds\frac{1}{2}|\clL_{k,\chi(k)}(\pmb{W}^{(\kappa)})|^2(\alpha_{k}/\alpha^{(\kappa)}_{k}
+\beta_{\chi(k)}/\beta^{(\kappa)}_{\chi(k)})
]^{-1} \label{ap1}
\end{eqnarray}
over the trust region
\begin{equation}\label{tru1}
2\Re\{\clL_{k,\chi(k)}(\pmb{W})(\clL_{k,\chi(k)}(\pmb{W}^{(\kappa)}))^*\}
-\frac{1}{2}|\clL_{k,\chi(k)}(\pmb{W}^{(\kappa)})|^2(\alpha_{k}/\alpha^{(\kappa)}_{k}
+\beta_{\chi(k)}/\beta^{(\kappa)}_{\chi(k)})>0,
\end{equation}
for $x_{k,\chi(k)}^{(\kappa)}=|\clL_{k,\chi(k)}(\pmb{W}^{(\kappa)})|^2/\sqrt{\alpha^{(\kappa)}_{k}\beta^{(\kappa)}_{\chi(k)}}$,
\begin{equation}\label{ap3}
\begin{array}{c}
a_{k,\chi(k)}^{(\kappa)}=\ds\ln(1+x_{k,\chi(k)}^{(\kappa)})+\frac{x_{k,\chi(k)}^{(\kappa)}}{x_{k,\chi(k)}^{(\kappa)}+1}>0,\\
b_{k,\chi(k)}^{(\kappa)}=\ds\frac{(x_{k,\chi(k)}^{(\kappa)})^2}{x_{k,\chi(k)}^{(\kappa)}+1}>0.
\end{array}
\end{equation}
\end{myth}
\Prf We use the following inequalities with their proof given in the Appendix:
\begin{equation}\label{inequa1}
\ln(1+x) \geq\ds \ln(1+\bar{x}) {\color{black}+} \frac{\bar{x}}{\bar{x}+1}
-\frac{\bar{x}^2}{\bar{x}+1}\frac{1}{x}
\quad \forall x>0, \bar{x}>0
\end{equation}
and
\begin{equation}\label{inequa2}
\ds\frac{|x|^2}{\sqrt{\alpha\beta}}\geq 2\frac{\Re\{x\bar{x}^*\}}{\sqrt{\bar{\alpha}\bar{\beta}}}
-\frac{1}{2}\frac{|\bar{x}|^2}{\sqrt{\bar{\alpha}\bar{\beta}}}
(\alpha/\bar{\alpha}+\beta/\bar{\beta})\quad \forall x\in\mathbb{C}, \bar{x}\in\mathbb{C},\alpha>0,
\bar{\alpha}>0, \beta>0,
\bar{\beta}>0,
\end{equation}
which is the same as
\begin{equation}\label{inequa3}
\ds\frac{\sqrt{\alpha\beta}}{|x|^2}\leq \frac{\sqrt{\bar{\alpha}\bar{\beta}}}{2\Re\{x\bar{x}^*\}
-\frac{1}{2}|\bar{x}|^2(\alpha/\bar{\alpha}+\beta/\bar{\beta})}\quad \forall x\in\mathbb{C}, \bar{x}\in\mathbb{C},\alpha>0,
\bar{\alpha}>0, \beta>0,
\bar{\beta}>0.
\end{equation}
Applying (\ref{inequa1}) and (\ref{inequa3})
 for $x=|\clL_{k,\chi(k)}(\pmb{W})|^2/\sqrt{\alpha_{k}\beta_{\chi(k)}}$ and $\bar{x}=|\clL_{k,\chi(k)}(\pmb{W}^{(\kappa)})|^2/\sqrt{\alpha_{k}^{(\kappa)}\beta_{\chi(k)}^{(\kappa)}}$
yields
\[
\begin{array}{lll}
\ds\ln (1+\frac{|\clL_{k,\chi(k)}(\pmb{W})|^2}
{\sqrt{\alpha_{k}\beta_{\chi(k)}}})&\geq&a_{k,\chi(k)}^{(\kappa)}-b_{k,\chi(k)}^{(\kappa)}
\ds\frac{\sqrt{\alpha_{k}\beta_{\chi(k)}}}{|\clL_{k,\chi(k)}(\pmb{W})|^2}\\
&\geq&a_{k,\chi(k)}^{(\kappa)}-b_{k,\chi(k)}^{(\kappa)}{\color{black}\sqrt{\alpha_{k}^{(\kappa)}\beta_{\chi(k)}^{(\kappa)}}}
[2\Re\{\clL_{k,\chi(k)}(\pmb{W})(\clL_{k,\chi(k)}(\pmb{W}^{(\kappa)}))^*\}\\
&&-\ds\frac{1}{2}|\clL_{k,\chi(k)}(\pmb{W}^{(\kappa)})|^2(\alpha_{k}/\alpha^{(\kappa)}_{k}
+\beta_{\chi(k)}/\beta^{(\kappa)}_{\chi(k)})
]^{-1},
\end{array}
\]
showing (\ref{ap1}). \qed

Accordingly, for a feasible $(\pmb{W}^{(\kappa)}, \pmb{\alpha}^{(\kappa)}, \pmb{\beta}^{(\kappa)})$ of (\ref{MaxMinPair3})
found at the $(\kappa-1)$th iteration, the following convex optimization problem is solved
at the $\kappa$th iteration to generate the next feasible  $(\pmb{W}^{(\kappa+1)}, \pmb{\alpha}^{(\kappa+1)}, \pmb{\beta}^{(\kappa+1)})$:
\begin{equation}\label{ConvexOpt}
\begin{array}{r}
\ds\max_{\pmb{W}\in\mathbb{C}^{N\times N}, \pmb{\alpha}\in \mathbb{R}_+^{2K}, \pmb{\beta}\in \mathbb{R}_+^{2K}}
f^{(\kappa)}(\pmb{W},\pmb{\alpha},
\pmb{\beta})\triangleq \ds \min_{k=1,\ldots ,K}\ \frac{1}{r_k}
[f^{(\kappa)}_{k,K+k}(\pmb{W},\alpha_{k},\beta_{K+k})+f^{(\kappa)}_{K+k,k}(\pmb{W},\alpha_{K+k},\beta_{k})]\\
\mbox{subject to}\quad (\ref{DemoniCon3})-(\ref{TotalRelayCon3} ), (\ref{tru1}).\end{array}
\end{equation}
Note that
\[
f(\pmb{W}^{(\kappa+1)},\pmb{\alpha}^{(\kappa+1)},\pmb{\beta}^{(\kappa+1)})\geq f^{(\kappa)}(\pmb{W}^{(\kappa+1)},\pmb{\alpha}^{(\kappa+1)},
\pmb{\beta}^{(\kappa+1)})
\]
by (\ref{ap1}), and
\[
f(\pmb{W}^{(\kappa)},\pmb{\alpha}^{(\kappa)},\pmb{\beta}^{(\kappa)})= f^{(\kappa)}(\pmb{W}^{(\kappa)},\pmb{\alpha}^{(\kappa)},
\pmb{\beta}^{(\kappa)})
\]
because
\[
\ln (1+\frac{|\clL_{k,\chi(k)}(\pmb{W}^{(\kappa)})|^2}
{\sqrt{\alpha_{k}^{(\kappa)}\beta_{\chi(k)}^{(\kappa)}}})=f^{(\kappa)}_{k,\chi(k)}(\pmb{W}^{(\kappa)},\alpha_{k}^{(\kappa)},
\beta_{\chi(k)}^{(\kappa)}), k=1,\dots, 2K.
\]
On the other hand, as $(\pmb{W}^{(\kappa)},\pmb{\alpha}^{(\kappa)},\pmb{\beta}^{(\kappa)})$ and
$(\pmb{W}^{(\kappa+1)},\pmb{\alpha}^{(\kappa+1)},\pmb{\beta}^{(\kappa+1)})$ are a feasible point and the optimal
solution of the convex optimization problem (\ref{ConvexOpt}), it is true that
\[
f^{(\kappa)}(\pmb{W}^{(\kappa+1)},\pmb{\alpha}^{(\kappa+1)},
\pmb{\beta}^{(\kappa+1)})>f^{(\kappa)}(\pmb{W}^{(\kappa)},\pmb{\alpha}^{(\kappa)},
\pmb{\beta}^{(\kappa)})
\]
as far as $(\pmb{W}^{(\kappa)},\pmb{\alpha}^{(\kappa)},\pmb{\beta}^{(\kappa)})\neq
(\pmb{W}^{(\kappa+1)},\pmb{\alpha}^{(\kappa+1)},\pmb{\beta}^{(\kappa+1)})$. The point $(\pmb{W}^{(\kappa+1)},\pmb{\alpha}^{(\kappa+1)},\pmb{\beta}^{(\kappa+1)})$ is then better than $(\pmb{W}^{(\kappa)},\pmb{\alpha}^{(\kappa)},\pmb{\beta}^{(\kappa)})$ because
\[
f(\pmb{W}^{(\kappa+1)},\pmb{\alpha}^{(\kappa+1)},\pmb{\beta}^{(\kappa+1)})\geq f^{(\kappa)}(\pmb{W}^{(\kappa+1)},\pmb{\alpha}^{(\kappa+1)}>f^{(\kappa)}(\pmb{W}^{(\kappa)},\pmb{\alpha}^{(\kappa)},
\pmb{\beta}^{(\kappa)})=f(\pmb{W}^{(\kappa)},\pmb{\alpha}^{(\kappa)},\pmb{\beta}^{(\kappa)}).
\]
Analogously to \cite[Proposition 1]{TTN16}, it can be shown that the sequence $\{ (\pmb{W}^{(\kappa)},\pmb{\alpha}^{(\kappa)},\pmb{\beta}^{(\kappa)}) \}$ at least converges to a locally
optimal solution of the exchange throughput optimization problem (\ref{MaxMinPair3}). The proposed Algorithm \ref{alg1}
for (\ref{MaxMinPair3}) thus terminates after finitely many iteration, yielding an optimal solution
$(\mathbf{W}^{opt},\pmb{\alpha}^{opt},\pmb{\beta}^{opt})$
within tolerance $\epsilon>0$. Then $(\mathbf{W}^{opt},\pmb{p}^{opt})$ with $\pmb{p}^{opt}=
(1/\sqrt{\beta^{opt}_1},\dots,1/\sqrt{\beta^{opt}_{2K}})^T$
is accepted as the computational solution of the maximin
exchange throughput optimization problem (\ref{MaxMinPair2}).
\begin{algorithm}
\caption{Path-following algorithm for exchange throughput optimization} \label{alg1}
\begin{algorithmic}
\STATE \textbf{initialization}: Set $\kappa=0$. Choose an initial feasible point
$(\pmb{W}^{(0)}, \pmb{\alpha}^{(0)}, \pmb{\beta}^{(0)})$ for the convex constraints (\ref{DemoniCon3})-(\ref{TotalRelayCon3} ).
Calculate $R_{\min}^{(0)}=\ds{\min_{k=1,\ldots ,K}}
R_k(\pmb{W}^{(0)}, \pmb{\alpha}^{(0)}, \pmb{\beta}^{(0)})$.
\REPEAT \STATE $\bullet$ Set
$\kappa=\kappa+1$.
\STATE $\bullet$ Solve the
convex optimization problem \eqref{ConvexOpt} to obtain the
solution $(\pmb{W}^{(\kappa)}, \pmb{\alpha}^{(\kappa)}, \pmb{\beta}^{(\kappa)})$.
\STATE $\bullet$ Calculate $R_{\min}^{(\kappa)}=\ds{\min_{k=1,\ldots ,K}} R_k(\pmb{W}^{(\kappa)}, \pmb{\alpha}^{(\kappa)}, \pmb{\beta}^{(\kappa)})$.
 \UNTIL{$\frac{R_{\min}^{(\kappa)}-R_{\min}^{(\kappa-1)})}{R_{\min}^{(\kappa-1)}} \leq
 \epsilon$}.
\end{algorithmic}
\end{algorithm}

Before closing this section,  it is pointed out that the one-way
relay optimization in which UE $k$ sends information to
UE  $K+k$  can be formulated as in (\ref{MaxMinPair3}) by
setting $\gamma_{k}=0$ and  $p_{K+k}=0$ and thus can be directly solved by Algorithm \ref{alg1}.
\section{Energy efficiency maximization}\label{sec:ee}
Return to consider the EE maximization problem (\ref{e1}). It is worth noticing that the computational solution for the
QoS constrained sum throughput maximization problem
\begin{eqnarray}\label{sume1}
\ds\max_{\pmb{W}\in\mathbb{C}^{N\times N}, \pmb{p}\in \mathbb{R}_+^{2K}}   \sum_{k=1}^K\left[\ln (1+ \gamma_{k}(\pmb{p}, \pmb{W}))+\ln (1+\gamma_{K+k}(\pmb{p}, \pmb{W}))\right]\nonumber\\
\mbox{subject to}\quad
\eqref{IndiUserCon}, \eqref{SumUserCon}, \eqref{IndiRelayCon}, \eqref{SumRelayCon}, (\ref{e1c}),
\end{eqnarray}
which is a particular case of (\ref{e1}), is still largely open. As a by-product, our approach to computation for (\ref{e1})
is directly applicable to that for (\ref{sume1}).
Similarly to Theorem \ref{basic}, it can be shown that (\ref{e1}) is equivalently expressed by the following optimization
problem under the variable change (\ref{betak}):
\begin{subequations}\label{e1.eq}
\begin{eqnarray}
\ds\max_{\pmb{W}\in\mathbb{C}^{N\times N}, \pmb{\alpha}\in \mathbb{R}_+^{2K}, \pmb{\beta}\in \mathbb{R}_+^{2K}} F(\pmb{W},\pmb{\alpha},
\pmb{\beta})\triangleq\ds\frac{\sum_{k=1}^K
[\ln (1+\frac{|\clL_{k,K+k}(\pmb{W})|^2}
{\sqrt{\alpha_{k}\beta_{K+k}}})+\ln (1+\ds\frac{|\clL_{K+k,k}(\pmb{W})|^2}
{\sqrt{\alpha_{K+k}\beta_{k}}})]}{\pi(\pmb{\beta},\pmb{W})} \label{e1.eqa}\\
\mbox{subject to} \quad (\ref{DemoniCon3})-(\ref{TotalRelayCon3}),\label{eq.eqb}\\
\tilde{R}_k(\mathbf{W},\pmb{\alpha},\pmb{\beta})\triangleq \ln (1+\frac{|\clL_{k,K+k}(\pmb{W})|^2}
{\sqrt{\alpha_{k}\beta_{K+k}}})+\ln (1+\ds\frac{|\clL_{K+k,k}(\pmb{W})|^2}
{\sqrt{\alpha_{K+k}\beta_{k}}})\geq r_k, k=1,\dots, K,\label{e1.eqc}
\end{eqnarray}
\end{subequations}
where  the consumption power function $\pi(\pmb{\beta},\pmb{W})$ is defined by
\begin{equation}\label{e2}
\pi(\pmb{\beta},\pmb{W})\triangleq \sum_{k=1}^{2K}\frac{\zeta}{\sqrt{\beta_k}}+\zeta\ds\sum_{m=1}^M[\ds \sum_{\ell=1}^{2K}\Phi_{\ell,m}(\pmb{W}_m,1,\beta_{\ell}) +\sigma_R^2||\pmb{W}_m||^2 ]+MP^{\rm R}+2KP^{\rm U}.
\end{equation}
In Dinkelbach's iteration based approach (see e.g. \cite{Buetal16}), one aims to find through bisection
a value $\tau$ such that the optimal value of the following optimization problem is zero
\begin{eqnarray}
\ds\max_{\pmb{W}\in\mathbb{C}^{N\times N}, \pmb{\alpha}\in \mathbb{R}_+^{2K}, \pmb{\beta}\in \mathbb{R}_+^{2K}} \
\left[\ds\sum_{k=1}^K(\ln (1+\frac{|\clL_{k,K+k}(\pmb{W})|^2}
{\sqrt{\alpha_{k}\beta_{K+k}}})+\ln (1+\ds\frac{|\clL_{K+k,k}(\pmb{W})|^2}
{\sqrt{\alpha_{K+k}\beta_{k}}}))\right.\nonumber\\
\left.- \pi(\pmb{\beta},\pmb{W})\right] \quad
\mbox{subject to} \quad (\ref{DemoniCon3})-(\ref{TotalRelayCon3}). (\ref{e1.eqc}).\label{ditert}
\end{eqnarray}
Such $\tau$ obviously is the optimal value of (\ref{e1.eq}). However, for each $\tau$,  (\ref{ditert}) is still
nonconvex and as computationally difficult as the original optimization problem (\ref{e1.eq}).
There is no benefit to use (\ref{ditert}).

To address computation for (\ref{e1.eq}) involving the  nonconcave objective function $F(\pmb{W},\pmb{\alpha},
\pmb{\beta})$ and the nonconvex constraint (\ref{e1.eqc}),
we  will explore the following inequality for positive quantities, whose proof is given
in the Appendix:
\begin{eqnarray}
\ds\frac{\ln(1+x)}{t}&\geq& 2\ds\frac{\ln(1+\bar{x})}{\bar{t}}+\frac{\bar{x}}{\bar{t}(\bar{x}+1)}-\frac{\bar{x}^2}{(\bar{x}+1)\bar{t}}\frac{1}{x}-
\frac{\ln(1+\bar{x})}{\bar{t}^2}t \label{ine1}\\
&&\quad\forall\ x>0, \bar{x}>0, t>0, \bar{t}>0.\nonumber
\end{eqnarray}
The right-hand-side (RHS) of (\ref{ine1}) is a concave function on the interior domain of ${\cal R}^2_+$
and agrees with the left-hand-side (LHS) at $(\bar{x},\bar{t})$.

Suppose that $(\pmb{W}^{(\kappa)},\pmb{\alpha}^{(\kappa)},\pmb{\beta}^{(\kappa)})$ is a feasible point of (\ref{e1.eq})
found from the $(\kappa-1)$th iteration. Applying (\ref{ine1}) for
\[
x=|\clL_{k,\chi(k)}(\pmb{W})|^2/\sqrt{\alpha_{k}\beta_{\chi(k)}}, t=\pi(\pmb{\beta},\pmb{W})
\]
and
\[
\bar{x}=|\clL_{k,\chi(k)}(\pmb{W}^{(\kappa)})|^2/\sqrt{\alpha_{k}^{(\kappa)}\beta_{\chi(k)}^{(\kappa)}},
\bar{t}=\pi(\pmb{\beta}^{(\kappa)},\pmb{W}^{(\kappa)})
\]
yields
\begin{eqnarray}
\ds\frac{\ln (1+\ds\frac{|\clL_{k,\chi(k)}(\pmb{W})|^2}
{\sqrt{\alpha_{k}\beta_{\chi(k)}}})}{\pi(\pmb{\beta},\pmb{W})}&\geq&F_{k,\chi(k)}^{(\kappa)}(\pmb{W},\alpha_{k},
\pmb{\beta})\nonumber\\
&\triangleq&p_{k,\chi(k)}^{(\kappa)}- {\color{black}q_{k,\chi(k)}^{(\kappa)}\sqrt{\alpha^{(\kappa)}_{k}\beta^{(\kappa)}_{\chi(k)} }}[2\Re\{\clL_{k,\chi(k)}(\pmb{W})(\clL_{k,\chi(k)}(\pmb{W}^{(\kappa)}))^*\} \nonumber\\
&&-\ds\frac{1}{2}|\clL_{k,\chi(k)}(\pmb{W}^{(\kappa)})|^2(\alpha_{k}/\alpha^{(\kappa)}_{k}
+\beta_{\chi(k)}/\beta^{(\kappa)}_{\chi(k)})
]^{-1}\nonumber\\
&&-r_{k,\chi(k)}^{(\kappa)}\pi(\pmb{\beta},\pmb{W}),\label{e7}
\end{eqnarray}
over the trust region (\ref{tru1}),  for $x_{k,\chi(k)}^{(\kappa)}=|\clL_{k,\chi(k)}(\pmb{W}^{(\kappa)})|^2/\sqrt{\alpha^{(\kappa)}_{k}\beta^{(\kappa)}_{\chi(k)}}$,
\begin{equation}\label{e8}
\begin{array}{lll}
t^{(\kappa)}&=&\pi(\pmb{\beta}^{(\kappa)},\pmb{W}^{(\kappa)}),\\
p_{k,\chi(k)}^{(\kappa)}&=&2\ds\frac{\ln(1+x_{k,\chi(k)}^{(\kappa)})}{t^{(\kappa)}}+
\frac{x_{k,\chi(k)}^{(\kappa)}}{t^{(\kappa)}(x_{k,\chi(k)}^{(\kappa)}+1)}>0\\
q_{k,\chi(k)}^{(\kappa)}&=&\ds\frac{(x_{k,\chi(k)}^{(\kappa)})^2}{(x_{k,\chi(k)}^{(\kappa)}+1)t^{(\kappa)}}>0, \\
r_{k,\chi(k)}^{(\kappa)}&=&\ds\frac{\ln(1+x_{k,\chi(k)}^{(\kappa)})}{(t^{(\kappa)})^2}>0.
\end{array}
\end{equation}
Accordingly, the following convex optimization problem is solved at the $\kappa$th iteration  to generate the next
iterative point $(\pmb{W}^{(\kappa+1)},\pmb{\alpha}^{(\kappa+1)},\pmb{\beta}^{(\kappa+1)})$:
\begin{subequations}\label{ConvexOpt.e}
\begin{eqnarray}
\ds\max_{\pmb{W}\in\mathbb{C}^{N\times N}, \pmb{\alpha}\in \mathbb{R}_+^{2K}, \pmb{\beta}\in \mathbb{R}_+^{2K}} F(\pmb{W},\pmb{\alpha},
\pmb{\beta})\triangleq
\ds {\color{black}
\sum_{k=1}^K}
[F^{(\kappa)}_{k,K+k}(\pmb{W},\alpha_{k},\pmb{\beta})+F^{(\kappa)}_{K+k,k}(\pmb{W},\alpha_{K+k},\pmb{\beta})]\label{opt.a}\\
\mbox{subject to}\quad (\ref{DemoniCon3})-(\ref{TotalRelayCon3} ), (\ref{tru1}), (\ref{e2}),\label{opt.b}\\
f^{(\kappa)}_{k,K+k}(\pmb{W},\alpha_{k},\beta_{K+k}) +f^{(\kappa)}_{K+k,k}(\pmb{W},\alpha_{K+k},\beta_{k})
\geq r_k, k=1,\dots, K,\label{opt.c}
\end{eqnarray}
\end{subequations}
where $f^{(\kappa)}_{k,\chi(k)}$ are defined from (\ref{ap1}). Since,
\[
f^{(\kappa)}_{k,K+k}(\pmb{W},\alpha_{k},\beta_{K+k}) +f^{(\kappa)}_{K+k,k}(\pmb{W},\alpha_{K+k},\beta_{k})\leq
\tilde{R}_k(\mathbf{W},\pmb{\alpha},\pmb{\beta})
\]
it follows that the feasibility of the nonconvex constraint (\ref{e1.eqc}) in (\ref{e1.eq}) is guaranteed by that
of (\ref{opt.c}). Also,
\[
f^{(\kappa)}_{k,K+k}(\pmb{W}^{(\kappa)},\alpha_{k}^{(\kappa)},\beta_{K+k}^{(\kappa)}) +f^{(\kappa)}_{K+k,k}(\pmb{W}^{(\kappa)},\alpha_{K+k}^{(\kappa)},\beta_{k}^{(\kappa)})=
\tilde{R}_k((\pmb{W}^{(\kappa)},\pmb{\alpha}^{(\kappa)},\pmb{\beta}^{(\kappa)}))\geq r_{\min}
\]
because $(\pmb{W}^{(\kappa)},\pmb{\alpha}^{(\kappa)},\pmb{\beta}^{(\kappa)})$ is feasible for (\ref{e1.eq}) and thus
feasible for (\ref{e1.eqc}). Therefore, the convex optimization problem (\ref{ConvexOpt.e}) is always feasible. Analogously
to the previous section, the sequence $\{(\pmb{W}^{(\kappa)},\pmb{\alpha}^{(\kappa)},\pmb{\beta}^{(\kappa)})\}$ is seen
convergent at least to a locally optimal solution of problem (\ref{e1.eq}) and as thus the proposed
Algorithm \label{alg2} terminates after finitely many iteration, yielding an optimal solution
$(\pmb{W}^{(opt)},\pmb{\alpha}^{opt},\pmb{\beta}^{opt})$ within tolerance $\epsilon$. Then
$(\pmb{W}^{opt},\pmb{p}^{opt})$ with $\pmb{p}^{opt}=(1/\sqrt{\beta_1^{opt}},...,1/\sqrt{\beta_{2K}^{opt}})^T$ is accepted as
the computational solution of the EE maximization problem (\ref{e1}).
\begin{algorithm}
\caption{Path-following Algorithm for Energy Efficiency } \label{alg2}
\begin{algorithmic}
\STATE \textbf{initialization}: Set $\kappa=0$. Choose an initial feasible point
$(\pmb{W}^{(0)}, \pmb{\alpha}^{(0)}, \pmb{\beta}^{(0)})$  and calculate
$e_{\max}^{(0)}$ as the value of the objective in (\ref{e1}) at $(\pmb{W}^{(0)}, \pmb{\alpha}^{(0)},
\pmb{\beta}^{(0)})$.
\REPEAT \STATE $\bullet$ Set
$\kappa=\kappa+1$.
\STATE $\bullet$ Solve the
convex optimization problem \eqref{ConvexOpt.e} to obtain the
solution $(\pmb{W}^{(\kappa)}, \pmb{\alpha}^{(\kappa)}, \pmb{\beta}^{(\kappa)})$.
\STATE $\bullet$ Calculate  $e_{\max}^{(\kappa)}$ as the value of the objective function in (\ref{e1}) at
$(\pmb{W}^{(\kappa)}, \pmb{\alpha}^{(\kappa)}, \pmb{\beta}^{(\kappa)})$.
 \UNTIL{$\frac{e_{\max}^{(\kappa)}-e_{\max}^{(\kappa-1)})}{e_{\max}^{(\kappa-1)}} \leq
 \epsilon$}.
\end{algorithmic}
\end{algorithm}

To find an initial feasible point $(\pmb{W}^{(0)}, \pmb{\alpha}^{(0)}, \pmb{\beta}^{(0)})$ for Algorithm \ref{alg2},
we use Algorithm \ref{alg1} for
computing (\ref{MaxMinPair2}), which terminates upon
\begin{equation}\label{inipoint}
[\ds{\min_{k=1,\ldots ,K}} R_k(\pmb{W}^{(\kappa)}, \pmb{\alpha}^{(\kappa)}, \pmb{\beta}^{(\kappa)})/r_k]\geq r_{\min}
\end{equation}
to satisfy (\ref{e1c}).

For the QoS constraints (\ref{indi2}) instead of (\ref{e1c}), which by the variable change (\ref{betak})
are equivalent to the following constraints
\begin{equation}\label{indi3}
|\clL_{k,\chi(k)}(\pmb{W})|^2-(e^{\tilde{r}_{\min}}-1)\sqrt{\alpha_k\beta_{\chi(k)}}\geq 0, k=1,\dots, 2K.
\end{equation}
The LHS of (\ref{indi3}) is a convex functions, so (\ref{indi3}) is called reverse convex \cite{Tuybook}, which can
be easily innerly approximated by linear approximation of the LHS at $(\pmb{W}^{(\kappa)},\alpha_k^{(\kappa)},
\beta_{\chi(k)}^{(\kappa)})$.

{\bf Remark.} To compare the energy efficiency with one-way communication we need to revisit the one-way model: the users
$\{1,\cdots,K\}$ send symbols $(s_1,\cdots,s_K)^T\in\mathbb{C}^K$ via the relays in the first stage and the
users $\{K+1,...,2K\}$ send symbols $(s_{K+1},\cdots,s_{2K})^T\in\mathbb{C}^K$ via the relays in the second state.
Denote by $\pmb{W}^1_m$ and $\pmb{W}^2_m$ the beamforming matrices for the received signals from the users
$\{1,\cdots,K\}$ and $\{K+1,\cdots,2K\}$, respectively.  The transmit power at relay $m$ in forwarding signals to users $\{K+1,\cdots, 2K\}$ in the first stage is
\[
  P_{m}^{A,1}(\pmb{p}^1,\pmb{W}^1_m) = \ds \sum_{\ell=1}^{K} p_{\ell} ||\pmb{W}^1_m\pmb{h}_{\ell,m}||^2+
   \sigma_R^2||\pmb{\pmb{W}^1}_m||^2,
\]
and the transmit power at relay $m$ in forwarding signals to users $\{1,\cdots, K\}$ in the second stage is
\[
\begin{array}{lll}
  P_{m}^{A,2}(\pmb{p}^2,\pmb{W}^2_m) = \ds \sum_{\ell=1}^{K} p_{\ell+K} ||\pmb{W}^2_m\pmb{h}_{\ell,m}||^2+
   \sigma_R^2||\pmb{W}^2_m||^2.
\end{array}
\]
Therefore, the power constraint  at  relay $m$ is
\begin{eqnarray}
\ds \sum_{\ell=1}^{K}
 ( p_{\ell}||\pmb{W}^1_m\pmb{h}_{\ell,m}||^2+ p_{\ell+K}||\pmb{W}^2_m\pmb{h}_{\ell+K,m}||^2)+
   \sigma_R^2(||\pmb{W}^1_m||^2+||\pmb{W}^2_m||^2)\leq P_m^{A,\max}, \label{one1}\\
   m=1,...,M.\nonumber
\end{eqnarray}
The total  power constraint is
\begin{eqnarray}
\ds\sum_{m=1}^M[\ds \sum_{\ell=1}^{K}
 ( p_{\ell}||\pmb{W}^1_m\pmb{h}_{\ell,m}||^2+ p_{\ell+K}||\pmb{W}^2_m\pmb{h}_{\ell+K,m}||^2)+
   \sigma_R^2(||\pmb{W}^1_m||^2+||\pmb{W}^2_m||^2)
]\leq  P_{\mathrm{sum}}^{R,\max}.\label{one2}
\end{eqnarray}
Accordingly, for $k=1,\cdots, K$, the SINR at UEs can be calculated as:
\begin{eqnarray}
\tilde{\gamma}_{K+k}(\pmb{p}^1, \pmb{W}^1)&=&\ds \frac{p_{k}\left|\ds\sum_{m=1}^M\pmb{f}_{m,K+k}^H
\pmb{W}^1_m\pmb{h}_{k,m}\right|^2}{
\ds\sum_{\ell=1, \ell \neq k}^{K} p_{\ell}\left|\sum_{m=1}^M\pmb{f}_{m,K+k}^H \pmb{W}^1_m\pmb{h}_{\ell,m}\right|^2
+\sigma_R^2\sum_{m=1}^M||\pmb{f}_{m,K+k}^H \pmb{W}^1_m||^2+\sigma^2_{K+k}}\nonumber\\
&=&\ds p_{k}|\clL_{K+k,k}(\pmb{W}^1)|^2/
[\ds\sum_{\ell=1, \ell \neq k}^{K} p_{\ell}|\clL_{K+k,\ell}(\pmb{W}^1)|^2\nonumber\\
&&+\sigma_R^2||\clL_{K+k}(\pmb{W}^1)||^2 +\sigma^2_{K+k}]
\label{one4}
\end{eqnarray}
and
\begin{eqnarray}
\tilde{\gamma}_{k}(\pmb{p}^2, \pmb{W}^2)&=&\ds \frac{p_{K+k}\left|\ds\sum_{m=1}^M\pmb{f}_{m,k}^H
\pmb{W}^2_m\pmb{h}_{K+k,m}\right|^2}{
\ds\sum_{\ell=1, \ell \neq k}^{K} p_{K+\ell}\left|\sum_{m=1}^M\pmb{f}_{m,k}^H \pmb{W}^2_m\pmb{h}_{K+\ell,m}\right|^2
+\sigma_R^2\sum_{m=1}^M||\pmb{f}_{m,k}^H \pmb{W}^2_m||^2+\sigma^2_{k}}\nonumber\\
&=&\ds p_{K+k}|\clL_{k,K+k}(\pmb{W}^2)|^2/
[\ds\sum_{\ell=1, \ell \neq k}^{K} p_{K+\ell}|\clL_{k,K+\ell}(\pmb{W}^2)|^2\nonumber\\
&&+\sigma_R^2||\clL_k(\pmb{W}^2)||^2 +\sigma^2_{k}].
\label{one3}
\end{eqnarray}
The EE maximization problem is then formulated as
\begin{subequations}\label{onee1}
\begin{eqnarray}
&\ds\max_{\pmb{W}^1\in\mathbb{C}^{N\times N}, \pmb{W}^2\in\mathbb{C}^{N\times N}, \pmb{p}\in \mathbb{R}_+^{2K}} & \ds  {\color{black} \frac{ \ds\frac{1}{2}\sum_{k=1}^K\left[\ln (1+ \tilde{\gamma}_{k}(\pmb{p}^2, \pmb{W}^2))+\ln (1+\tilde{\gamma}_{K+k}(\pmb{p}^1, \pmb{W}^1))\right]}{\zeta(P^U_{\rm sum}(\mathbf{p})+\tilde{P}^R_{\rm sum}(\mathbf{p},
\mathbf{W}))+2MP^{\rm R}+2KP^{\rm U}}} \label{onee1a}\\
&\mbox{subject to} & \eqref{IndiUserCon}, \eqref{SumUserCon}, (\ref{one1}), (\ref{one2}), \label{onee1b}\\
&&\ln (1+ \tilde{\gamma}_{k}(\pmb{p}^2, \pmb{W}^2))+\ln (1+\tilde{\gamma}_{K+k}(\pmb{p}^1, \pmb{W}^1))
\geq r_{\min}, \label{onee1c}\\
&&k=1,\cdots, K. \nonumber
\end{eqnarray}
\end{subequations}
The pre-log factor $1/2$ in the numerator of (\ref{onee1a}) is to account for two stages needed in
communicating $s_1,\cdots, 2K$ and the non transmission power consumption at the relays  $2MP^{\rm R}$ to
reflect the fact that the relays have to transmit twice.
\section{Numerical Results}\label{sec:Simulation}
This sections evaluates the proposed algorithms  through the simulation.
The channels in the receive signal equations (\ref{receivedr}) and (\ref{receivedu}) are assumed Rayleigh fading,
which are modelled by independent circularly-symmetric complex Gaussian random variables
with zero mean and unit variance, while  the background noises $\pmb{n}_{R,m}$
and $n_k$  are also normalized, i.e., $\sigma_R^2=\sigma_k^2=1$. The computation tolerance
for terminating the Algorithm is $\epsilon=10^{-4}$.
The numerical results are averaged over $1,000$ random channel realizations.

Without loss of generality, simply set $P_k^{U, \max}\equiv P^{U, \max}$, and $P_n^{A, \max}\equiv P^{A, \max}$, $P_{\mathrm{sum}}^{U,\max}=KP^{U, \max}$, and $P_{\mathrm{sum}}^{R,\max}=MP^{A, \max}/2$.
$P^{U, \max}$ is fixed at $10$ dBW  but the relay power budget $P_{\mathrm{sum}}^{R,\max}$
varies from $0$ to $30$ dBW.   {\color{black} The drain efficiency of power amplifier $1/\zeta$ is set
to be $40 \%$. As in \cite{Xiong2012}, the circuit powers of each antenna in relay and UE are $0.97$ dBW and $-13$ dBW, respectively.} We consider the scenarios of $K \in \{1,2,3\}$ pairs
and $(M,N_R) \in \{(1, 8), (2, 4), (4, 2)\}$, i.e. the total number of antennas is fixed at $8$ but the number of relays
is $M\in\{1, 2, 4\}$.

\subsection{Maximin exchange throughput optimization}
This subsection analyses  the exchange throughput achieved by  TWR.
The jointly optimal power and relay beamforming is referred  as OP-OW, while
the optimal beamforming weights with UE equal power allocation are referred as OW. The initial points for Algorithm
\ref{alg1} is chosen from the OW solutions.
To compare with the numerical result of \cite{Khetal13}, $r_k\equiv 1$ is set for (\ref{MaxMinPairObj2}).

\begin{figure}
\centering
\includegraphics[width=4.5in]{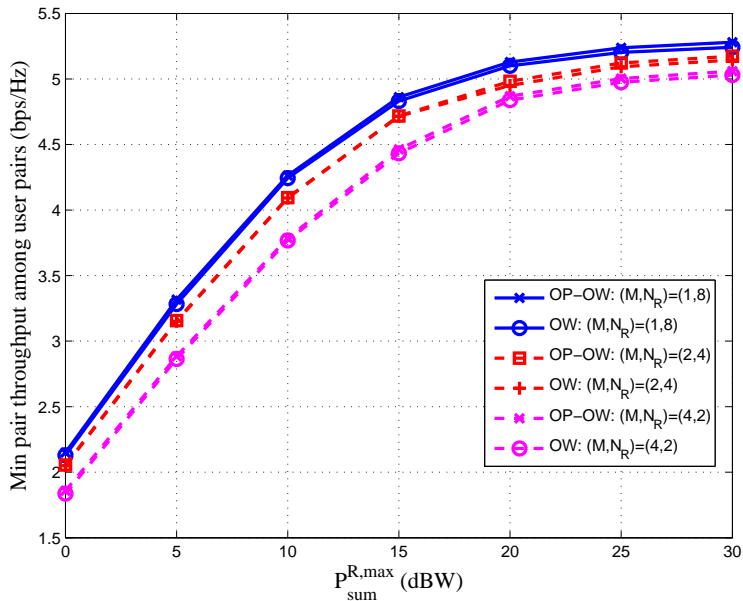}
\caption{Minimum pair throughput among user pairs versus the  relay power budget with $K=1$.}
\label{fig:Ex1_rate_K1}
\end{figure}

\begin{figure}
\centering
\includegraphics[width=4.5in]{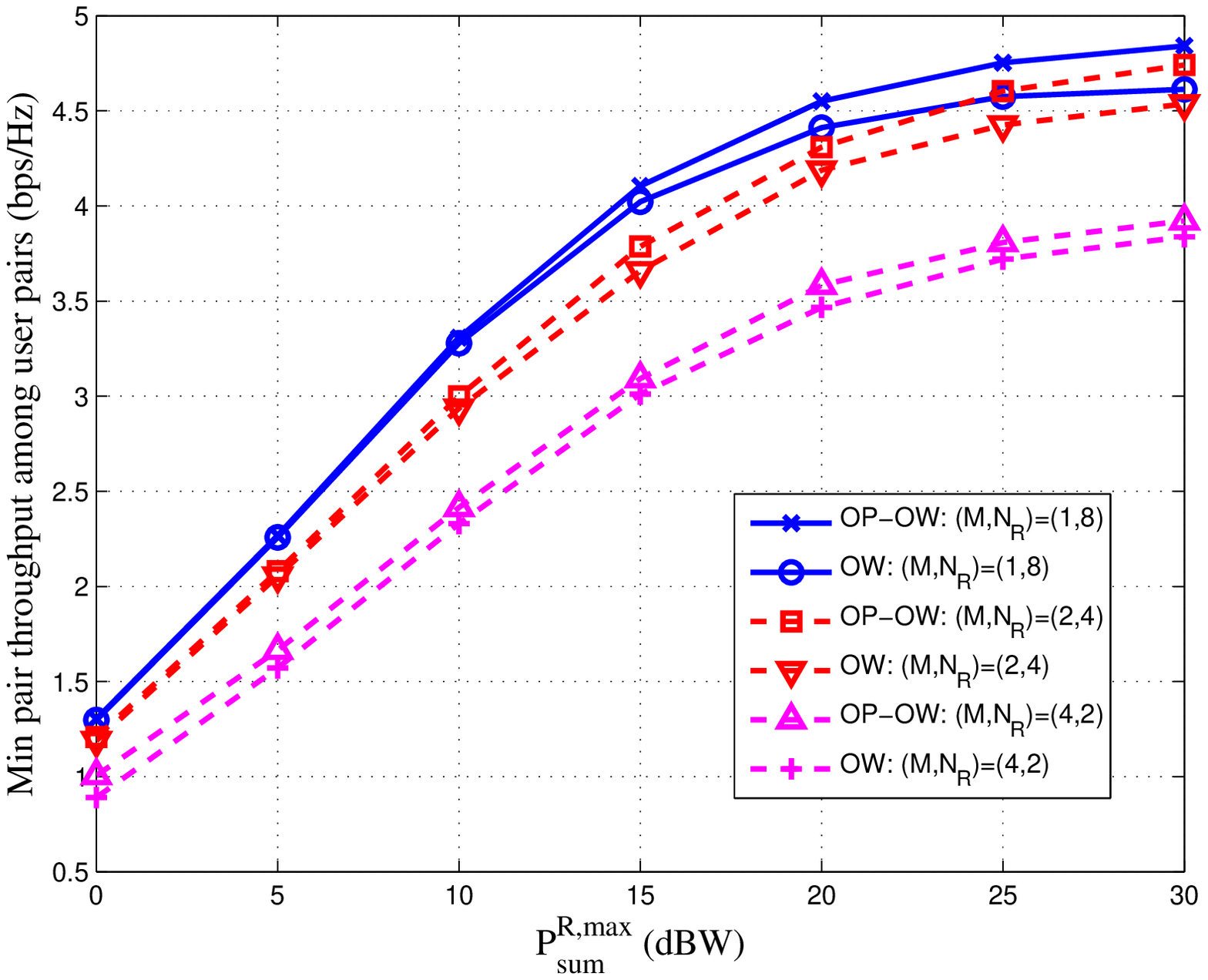}
\caption{Minimum pair throughput among user pairs versus the  relay power budget with $K=2$.}
\label{fig:Ex1_rate_K2}
\end{figure}

\begin{figure}
\centering
\includegraphics[width=4.5in]{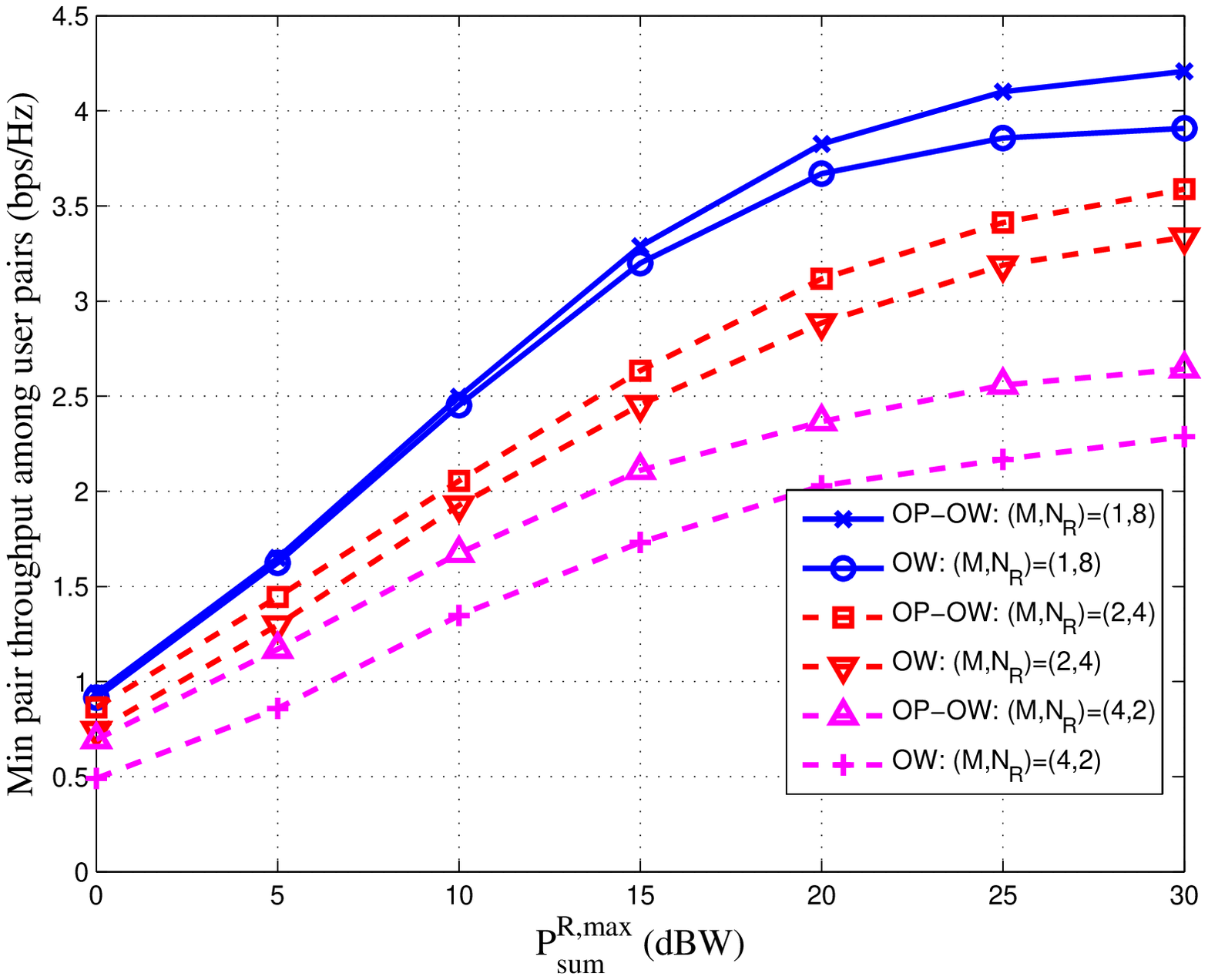}
\caption{Minimum pair throughput among user pairs versus the  relay power budget with $K=3$.}
\label{fig:Ex1_rate_K3}
\end{figure}

Fig. \ref{fig:Ex1_rate_K1}, \ref{fig:Ex1_rate_K2} and \ref{fig:Ex1_rate_K3} plot the
achievable  minimum pair exchange throughput versus the  relay power budget $P_{\mathrm{sum}}^{R,\max}$ with $K \in \{1,2,3\}$.
The improvement by OPOW over OW  is significant for $K=2$ and $K=3$.
The throughput gain is more considerable  with higher $P_{\mathrm{sum}}^{R,\max}$. It is also observed that
using less relays achieves better throughput.
 The result in Fig. \ref{fig:Ex1_rate_K2} for $K=2$ is consistent with that in \cite[Fig.2]{Khetal13}.

Table \ref{tab:Ex1_Ite} provides the average number of iterations of Algorithm \ref{alg1}. As can be observed,
Algorithm \ref{alg1}  converges in less than $24$ iterations in all considered scenarios.
\begin{table}[h!]
   \centering
   \caption{Average number of iterations of Algorithm \ref{alg1}  with $K=2$.}
   \begin{tabular}{ | c | c | c | c | c | c |  c |  c |}
    \hline
   $P_{\mathrm{sum}}^{R,\max}$ (dBW) & 0  &  5 &  10 &  15 &  20  &  25  & 30 \\
   \hline
   $(K, M,N_R)=(2,1,8)$ & 15.47 & 12.02 & 9.72 & 8.05 & 11.10 & 18.45 & 15.60 \\
    \hline
   $(K, M,N_R)=(2,2,4)$ & 24.20 & 11.80 & 8.47 & 7.07 & 10.70 & 11.22 & 13.17 \\
  \hline
  $(K, M,N_R)=(2,4,2)$ & 22.47 & 22.95 & 13.72 & 12.07 & 7.90 & 10.40 & 11.97 \\
  \hline
\end{tabular}
\label{tab:Ex1_Ite}
\end{table}

\subsection{EE maximization}
\begin{figure}
\centering
\includegraphics[width=4.5in]{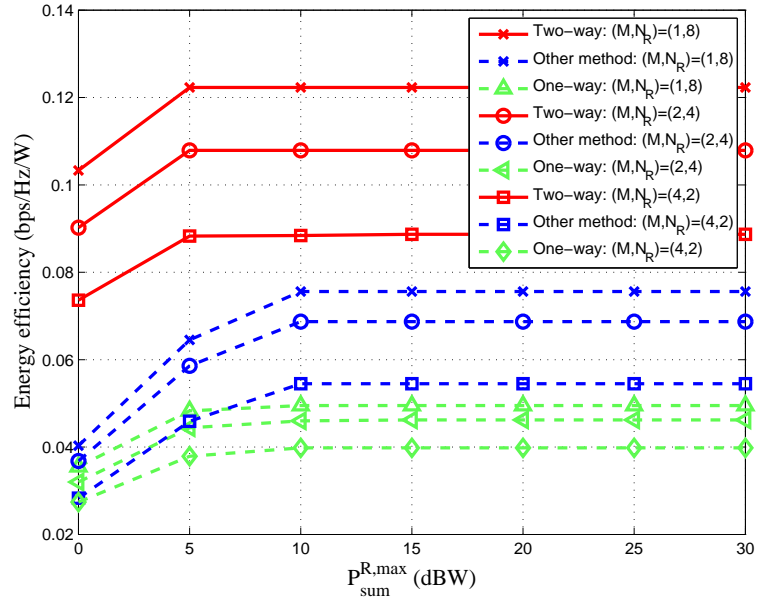}
\caption{Energy efficiency versus the  relay power budget with $K=2$}
\label{fig:Ex2_EE_K2}
\end{figure}

\begin{figure}
\centering
\includegraphics[width=4.5in]{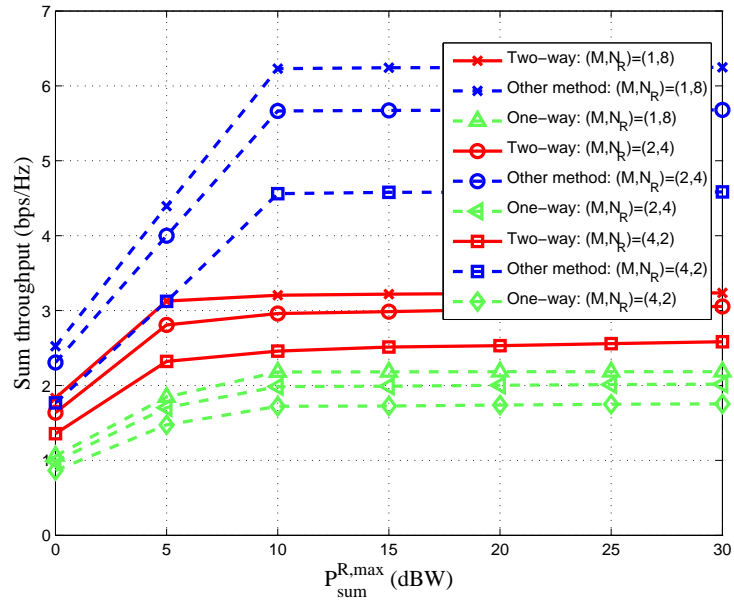}
\caption{Sum rate versus the relay power budget with $K=2$}
\label{fig:Ex2_Rate_K2}
\end{figure}

\begin{figure}
\centering
\includegraphics[width=4.5in]{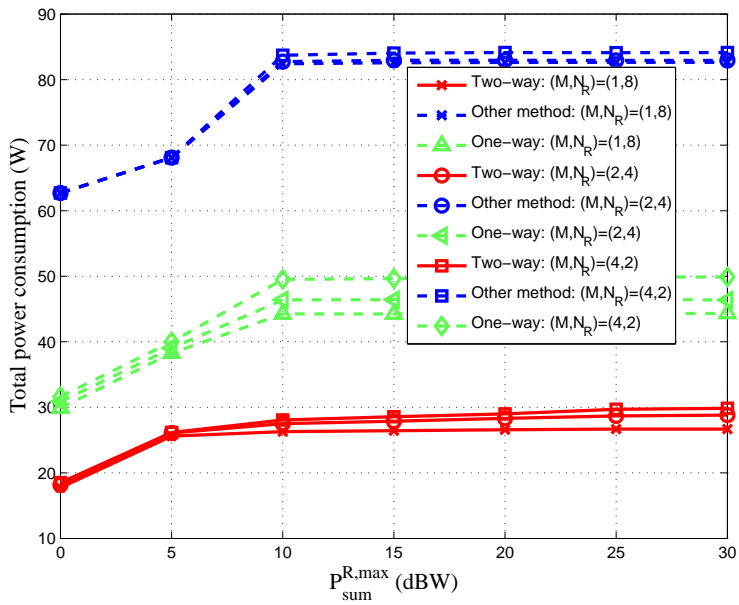}
\caption{Total power versus the relay power budget with $K=2$}
\label{fig:Ex2_Power_K2}
\end{figure}

This subsection examines the performance of energy efficiency achieved  by Algorithm \ref{alg2}.
$r_{k}$ in (\ref{inipoint}) is set as the half of the optimal value obtained by Algorithm \ref{alg1}.
Firstly, the simulation results presented in Fig. \ref{fig:Ex2_EE_K2}, \ref{fig:Ex2_Rate_K2} and \ref{fig:Ex2_Power_K2} are for
$K=2$ and $(M,N_R) \in \{(1, 8), (2, 4), (4, 2)\}$.
Fig. \ref{fig:Ex2_EE_K2} compares the EE performance achieved by TWR, one-way relaying
and TWR with UE fixed equal power allocation, which is labelled as "other method".
It is clear from Fig. \ref{fig:Ex2_EE_K2} that TWR clearly and significantly outperforms one-way relaying and
TWR UE equal power allocation.

Under small transmit power regime, the power consumption in the denominator is dominated by the circuit power and
the EE is maximized by maximizing the sum throughput in the numerator. As such all the EE, the sum
throughput and transmit power increase in Fig. \ref{fig:Ex2_EE_K2}, \ref{fig:Ex2_Rate_K2} and
\ref{fig:Ex2_Power_K2} in the relay power budget $P_{\mathrm{sum}}^{R,\max}$.

However, under larger transmit power regime, where the denominator of EE is dominated by the actual transmit power,
the EE becomes to be  maximized by minimizing the transmit power in the denominator,
which saturates when beyond a threshold. When the transmit power saturates in Fig. \ref{fig:Ex2_Power_K2}, both the
EE and the sum throughput accordingly saturate in Fig. \ref{fig:Ex2_EE_K2} and \ref{fig:Ex2_Rate_K2}.
It is also observed that for a given  relay power budget and a given number of total antennas in all relays, the configuration
with less relays is superior to the ones with more relays. This is quite expected since the configuration
with less relays achieves higher throughput than the ones with more relays.
Table \ref{tab:Ex2_Ite} shows that Algorithm \ref{alg2} converges
in less than $26$ iterations.

\begin{figure}
\centering
\includegraphics[width=4.5in]{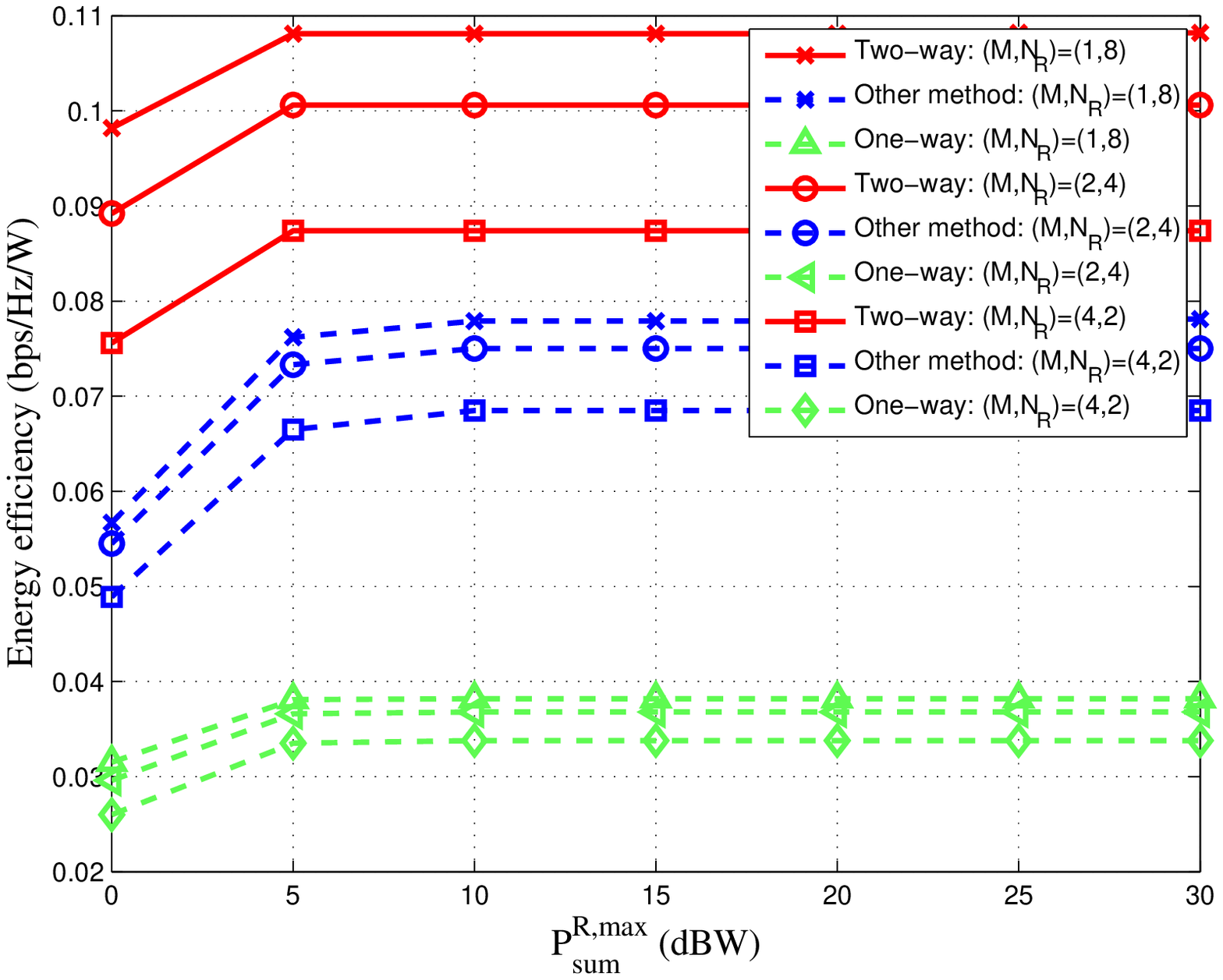}
\caption{Energy efficiency versus the  relay power budget with $K=1$}
\label{fig:Ex2_EE_K1}
\end{figure}

\begin{figure}
\centering
\includegraphics[width=4.5in]{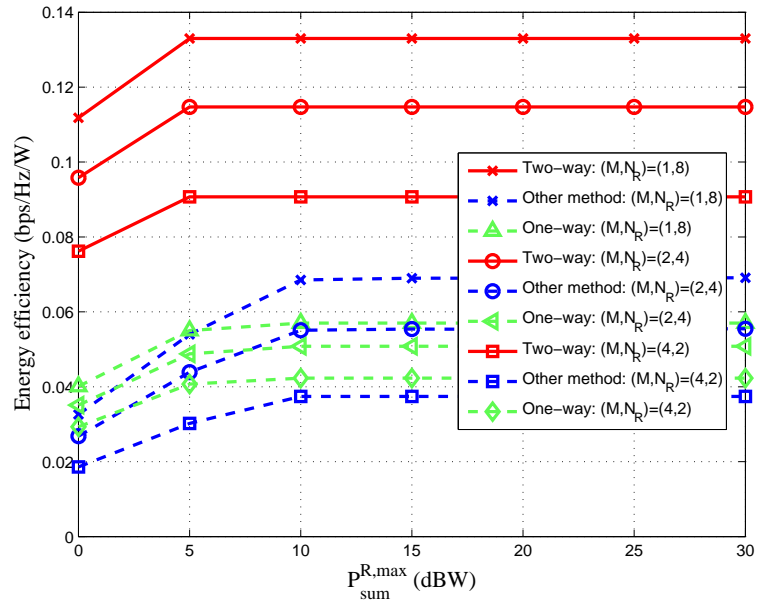}
\caption{Energy efficiency versus the relay power budget with $K=3$}
\label{fig:Ex2_EE_K3}
\end{figure}

Similar comparisons are provided in Fig. \ref{fig:Ex2_EE_K1} and \ref{fig:Ex2_EE_K3} for $K=1$ and $K=3$, respectively
with the superior  EE performances of TWR over one-way relaying and TWR UE equal power allocation observed.

\begin{table}[h!]
   \centering
   \caption{Average number of iterations of the proposed algorithm \ref{alg2} for two-way with $K=2$.}
   \begin{tabular}{ | c | c | c | c | c | c |  c |  c |}
    \hline
   $P_{\mathrm{sum}}^{R,\max}$ (dBW) & 0  &  5 &  10 &  15 &  20  &  25  & 30 \\
   \hline
   $(K, M,N_R)=(2,1,8)$ & 23.21 & 23.08 & 5.85 & 12.65 & 14.06 & 14.98 & 15.53 \\
    \hline
   $(K, M,N_R)=(2,2,4)$ & 20.78 & 26.05 & 6.85 & 13.75 & 19.73 & 19.98 & 20.45 \\
  \hline
  $(K, M,N_R)=(2,4,2)$ & 21.11 & 22.95 & 8.03 & 9.61 & 17.41 & 19.96 & 21.46 \\
  \hline
\end{tabular}
\label{tab:Ex2_Ite}
\end{table}

\section{Conclusions}\label{sec:Conclusion}
Joint UE power allocation and relay beamforming  to either satisfy a UE's QoS requirement or
maximize the energy efficiency of TWR serving multiple UEs is a very difficult nonconvex optimization problem.
This paper has developed two path-following computational procedures for their solutions, which invoke a simple
convex quadratic program of low computational complexity at each iteration.
Simulation results have confirmed their rapid convergence. We have shown that TWR achieves  much higher energy-efficiency
than its one-way relaying counterpart in all considered scenarios.
\section*{Appendix: proof for inequalities (\ref{inequa1}), (\ref{inequa2}) and (\ref{ine1})}
The function $\psi_1(z)=\ln (1+z^{-1})$ is convex on the domain $z>0$ \cite{Taetal16}. Therefore \cite{Tuybook}
\[
\psi_1(z)\geq \psi(\bar{z}) +\nabla\psi_1(\bar{z})(z-\bar{z})\ \quad\forall\ z>0, \bar{z}>0,
\]
which is seen as
\begin{equation}\label{apen1}
\ln (1+z^{-1})\geq \ln(1+\bar{z}^{-1})+\frac{1}{\bar{z}+1}-\frac{z}{(\bar{z}+1)\bar{z}}\quad\forall z>0, \bar{z}>0.
\end{equation}
The inequality (\ref{inequa1}) is obtained by substituting $x=z^{-1}$ and $\bar{x}=\bar{z}^{-1}$ into (\ref{apen1}).

Next, the function $\psi_2(x,\alpha,\beta)=|x|^2/\sqrt{\alpha\beta}$ is convex on the domain
$z\in\mathbb{C}, \alpha>0$ and $\beta>0$ \cite{DM08}, so again
\[
\psi_2(x,\alpha,\beta)\geq \psi_2(\bar{x},\bar{\alpha},\bar{\beta})+\la
\nabla \psi_2(\bar{x},\bar{\alpha},\bar{\beta}), (x,\alpha,\beta)-(\bar{x},\bar{\alpha},\bar{\beta})\ra,
\]
which is seen as (\ref{inequa2}).

Finally, by checking its Hessian, the function $\psi_3(z,t)=(\ln (1+z^{-1}))/t$ is seen to be
convex on the interior of ${\cal R}^2_+$.
Therefore,
\[
\psi_3(z,t)\geq \psi_2(\bar{z},\bar{t})+\la \nabla \psi_3(\bar{z},\bar{t}), (z,t)-(\bar{z},\bar{t})\ra \quad
\forall \ z>0, \bar{z}>0, t>0, \bar{t}>0,
\]
which is seen as
{\color{black}
\[
\frac{\ln(1+z^{-1})}{t}\geq 2\frac{\ln(1+\bar{z}^{-1})}{\bar{t}}+\frac{1}{(\bar{z}+1)\bar{t}}-
\frac{z}{(\bar{z}+1)\bar{z}\bar{t}}-\frac{\ln(1+\bar{z})}{\bar{t}^2}t.
\]}
The inequality (\ref{ine1}) follows by substituting $x=z^{-1}$ and $\bar{x}=\bar{z}^{-1}$ into the last inequality.

\end{document}